\documentclass[prl,twocolumn,showpacs,preprintnumbers,amsmath,amssymb,superscriptaddress,nofootinbib,english]{revtex4-1}

\usepackage{graphicx}
\usepackage{dcolumn}
\usepackage{bm}
\usepackage{epsfig}
\usepackage{graphicx}
\usepackage{hyperref}
\usepackage[usenames]{color}
\usepackage{url}

\hypersetup{
    colorlinks=true,
    linkcolor=red,
    citecolor=blue,
}

\newcommand{\tc}{\textcolor{black}}
\newcommand{\tr}{\textcolor{black}}
\newcommand{\remove}[1]{}

\def\etal{{\frenchspacing\it et al.}}
\def\ie{{\frenchspacing\it i.e.}}
\def\eg{{\frenchspacing\it e.g.}}

\def\be{\begin{equation}}
\def\ee{\end{equation}}
\def\ba{\begin{eqnarray}}
\def\ea{\end{eqnarray}}
\begin{document}

\title{Examining the evidence for dynamical dark energy}

\author{Gong-Bo Zhao}


\affiliation{Institute of Cosmology and Gravitation, University of Portsmouth,
Portsmouth, PO1 3FX, UK}

\affiliation{National Astronomy Observatories,
Chinese Academy of Science, Beijing, 100012, P.R.China}

\author{Robert G. Crittenden}
\affiliation{Institute of Cosmology and Gravitation, University of Portsmouth,
Portsmouth, PO1 3FX, UK}

\author{Levon Pogosian}
\affiliation{Department of Physics, Simon Fraser University, Burnaby, BC, V5A 1S6, Canada}
\affiliation{Institute of Cosmology and Gravitation, University of Portsmouth,
Portsmouth, PO1 3FX, UK}

\author{Xinmin Zhang}
\affiliation{Institute of High Energy Physics, Chinese Academy of Science (CAS), P.O. Box 918-4, Beijing 100049, P.R.China}
\affiliation{Theoretical Physics Center for Science Facilities (TPCSF), CAS, Beijing, 100049, P.R.China}

\begin{abstract}
We apply a new non-parametric Bayesian method for reconstructing the evolution history of the equation-of-state $w$ of dark energy, based on applying a correlated prior for $w(z)$, to a collection of cosmological data. We combine the latest supernova (SNLS 3-year or Union2.1), cosmic microwave background, redshift space distortion and the baryonic acoustic oscillation measurements (including BOSS, WiggleZ and 6dF) and find that the cosmological constant appears consistent with current data, but that a dynamical dark energy model which evolves from $w<-1$ at $z\sim0.25$ to  $w > -1$ at higher redshift is mildly favored. Estimates of the Bayesian evidences show little preference between the cosmological constant model and the dynamical model for a range of correlated prior choices.  Looking towards future data, we find that the best fit models for current data could be well distinguished from the $\Lambda$CDM model by observations such as Planck and Euclid-like surveys.  
\end{abstract}


\maketitle

Dark energy (DE), the source driving the acceleration of the universe in the framework of general relatively (GR), has remained an enigma since its discovery \cite{RiessPerl}. The accumulating observational data, including supernovae (SN), cosmic microwave background radiation (CMB) and large scale structure (LSS), constrains its equation-of-state $w(z)$, which can be a general function of redshift $z$. An accurate reconstruction of $w(z)$ can help us understand DE and gravity: a $w \ne -1$ would indicate a dynamical DE, while a $w(z)$ evolving across $-1$ would imply an additional intrinsic degree of freedom of DE and could be a smoking gun of the breakdown of Einstein's theory of general relativity on cosmological scales.   

The evolution of $w$ can be reconstructed from data using either parametric or non-parametric methods \cite{Sahni:2006pa}, with the latter having the advantage of not assuming any $ad~hoc$ form of $w(z)$. A common non-parametric approach is to bin $w$ in $z$, or 
the scale factor $a$, and fit the bin amplitudes to data. This assumes that $w(z)$ is constant within each bin, while the neighboring bins are treated as independent.  But it seems rather unphysical to assume 
perfect correlation of $w(z)$ within a bin, while having  
no correlation between different bins. This approach also leads to a practical problem --  when fitting many bins to data one finds that the uncertainties of binned $w$'s are very large and highly correlated, corresponding to flat directions in the likelihood function and a very slow convergence of Monte Carlo Markov chains (MCMC). Using only a few bins improves the convergence, but leads to unphysical discrete structures caused by the coarse binning. 

A principal component analysis (PCA) shows \cite{HutStark} that relatively few uncorrelated linear combinations of the bin amplitudes are well-constrained by data, while most are practically unconstrained. PCA provides a consistent framework for forecasting and comparing the information content of future surveys, as well as a tool for data compression~\cite{Crittenden:2005wj}. It has also been used to reconstruct $w(z)$ \cite{HutStark,Clarkson:2010bm} by first performing a Fisher forecast to determine the eigenmodes of the covariance matrix based on a particular fiducial model, and then fitting the few best constrained modes to data, while setting the amplitudes of the poorly constrained modes to zero. This simple truncation can lead to significant biases and unrealistically small errors in the reconstruction at high $z$, where the data are poor or absent entirely \cite{HutStark}. Ignoring the poorly constrained modes is equivalent to the unreasonable assumption that their amplitudes are measured to be zero with infinite accuracy.  In addition to biasing, zeroing the noisy modes introduces a hidden prior on the smoothness of $w(z)$ that is difficult to interpret and quantify. 

To address these 
issues, we recently introduced a new reconstruction method which imposes an explicit prior directly on the space of $w(z)$ functions, which is combined with observations in a straightforward Bayesian framework and is simple to implement \cite{Crittenden:2005wj,Crittenden:2011aa}. (The Gaussian Process method~\cite{Holsclaw:2010sk, Shafieloo:2012ht, Seikel:2012uu} is close in spirit to our approach, but the methods and interpretation of the results differ significantly.) The prior is independent of the choice of $w$ basis, and can be chosen to prefer smoothness in the reconstructions, naturally constraining the flat directions in parameter space which are most likely dominated by noise.  And while some bias in reconstructions is inevitable, particularly when one attempts to reconstruct $w(z)$ in regimes where the data are weak, by choosing a prior based on theoretically plausible $w(z)$,
we ensure that the biases are minimised for the most interesting models.  Here we apply this new method to a collection of the latest cosmological data to investigate whether there is evidence for dynamical dark energy.     

The choice of the prior is intrinsically subjective because it quantifies the initial probability distribution for the `true' model.  In principle, it should be determined by purely theoretical considerations, \eg~by considering ensembles of possible quintessence potentials, but here we take a more phenomenological approach.  We impose a smooth equation of state by assuming $w(a)$ is a Gaussian field with a covariance described by a correlation function 
\begin{equation}
 \xi_w (|a - a'|) \equiv \left\langle [w(a) - w^{\rm fid}(a)][w(a') - w^{\rm fid}(a')] \right\rangle \ ,
\end{equation}
which we take to be translationally invariant in $a$.  We assume the correlation function to have the form proposed by Crittenden, Pogosian and Zhao (CPZ)~\cite{Crittenden:2005wj}, 
\be
\label{eq:xi_CPZ} \xi_{\rm CPZ}(\delta a) =  \xi_w (0) /[1 + (\delta a/a_c)^2] \ , 
\ee 
where $a_c$ describes the typical smoothing distance, and $\xi_w(0)$ is a normalisation factor related to $\sigma^2_{\bar{w}}$, the amplitude of the expected variance of the mean $w$. 
The CPZ correlation function essentially requires that the $w$ bins within the correlation length $a_c$ are positively correlated, while the correlation falls off when $\delta{a}>a_c$. In \cite{Crittenden:2011aa} we have considered other choices of correlation functions and found that changing the shape of the correlation function does not significantly impact the reconstruction results. 

The two parameters, $a_c$ and $\xi_w(0)$ (or $\sigma^2_{\bar{w}}$) can be tuned to adjust the smoothing scale and the strength of the prior. 
A very weak prior (large $\sigma^2_{\bar{w}}$, small $a_c$) results in a noisy reconstruction (large variance), but which will match the true model on average (unbiased).  A stronger prior results in 
a smaller variance, but pulls the reconstructed model towards the peak of the prior, which will 
lead to a bias when the prior conflicts with the true model. In \cite{Crittenden:2011aa}, we considered reconstructing a range of models and found that, by setting $a_c=0.06$ and $\sigma_{\bar{w}}=0.02$, we were able to reconstruct them all without a sizeable bias.  However, as the functions we considered may not fully represent the full range of possible models, we also consider a weaker prior 
with a larger variance in the mean, $\sigma_{\bar{w}}=0.04$, which is still sufficient to curtail flat directions in the parameter space.  

\begin{figure*}[tbp]
\includegraphics[scale=0.3]{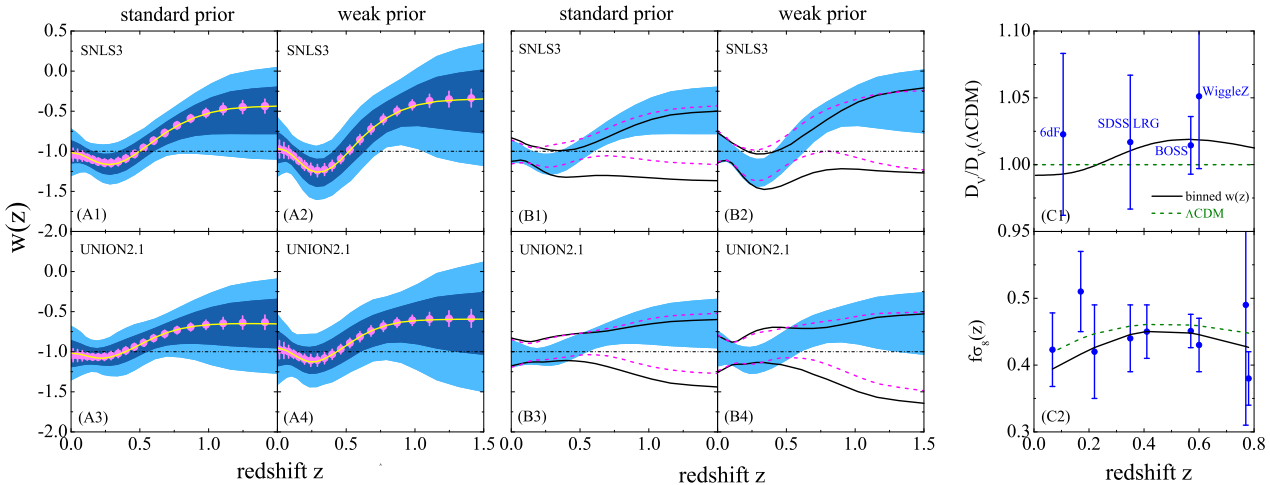}
\caption{Panels (A1-A4): The best fit $w(z)$ with marginalized 68 and 95\% CL error (shaded bands) reconstructed from a joint data set of the latest data including SN, CMB, $H(z)$, RSD and BAO. This best-fit $w(z)$ is used to construct mock data for future Planck and Euclid-like surveys, and the resulting reconstruction is shown by the magenta dots (with 68\% CL error bars.) The four panels show the result using different combinations of priors (standard and weak) and supernovae data sets (SNLS3 and Union2.1). Panels (B1-B4): The 68\% CL constraints on $w(z)$ using different data subsets: SN+CMB+$H(z)$(black solid), SN+CMB+$H(z)$+RSD(magenta dashed), SN+CMB+$H(z)$+RSD+BAO(blue band). Panels (C1,C2): These reconstructed models affect the fit to the BAO and RSD data; we compare the reconstruction from the combined data (using SNLS3 and weak prior) with the  $\Lambda$CDM model (green dashed). }\label{fig:wz}
\end{figure*}  

We 
parametrize $w(z)$ using a set of bins ${\bf w} \equiv \{w_i \}$, $i=1,...,N$, spaced uniformly in scale factor $a$ within 
$[a_{\rm min},1]$. 
To make $w(z)$ differentiable for the calculation of the dark energy perturbations, we interpolate between the bins using narrow {\tt tanh} functions \cite{Crittenden:2005wj}. As long as the bin width is small compared to the correlation length, the prior largely wipes out the dependence on the 
choice of binning.  The number of bins and the range are chosen to be large enough so that the results are stable to these choices, and alternative binning schemes, e.g. logarithmic in $z$, 
do not affect the results.  Given a binning scheme, it is straightforward to construct the prior bin covariance matrix ${\bf C}$~\cite{Crittenden:2005wj,Crittenden:2011aa}.  
The resulting prior in this space is ${\cal P}_{\rm prior}({\bf w}) \propto e^{-({\bf w}-{\bf w}^{\rm fid})^T{\bf C}^{-1}({\bf w}-{\bf w}^{\rm fid})/2}$, where ${\bf w}^{\rm fid}$ is the fiducial model. 

The reconstructed model is that which maximizes the posterior probability, which by Bayes' theorem is proportional to the likelihood of the data times the prior probability,  ${\cal P}({\bf w}|{\bf D}) \propto {\cal P}({\bf D}|{\bf w}) \times {\cal P}_{\rm prior}({\bf w})$.
Effectively, the prior results in a new contribution to the total $\chi^2$ of a model,  $\chi^2_{\rm prior} = ({\bf w}-{\bf w}^{\rm fid})^T{\bf C}^{-1}({\bf w}-{\bf w}^{\rm fid})$, which penalizes models that are less smooth. One can then use standard MCMC techniques to search for the minimum of $\chi^2=\chi^2_{\rm data} + \chi^2_{\rm prior}$. The prior curtails the flat directions in the likelihood corresponding to the unconstrained eigenmodes of ${\bf w}$, giving fast convergence of MCMC chains even with a large number of bins.

We wish to avoid explicit dependence on the fiducial model and there are several ways to marginalize over it~\cite{Crittenden:2011aa}. 
Here, we adopt the `floating' average method proposed in \cite{Crittenden:2011aa}, taking the fiducial model to be a flat local average of trial bin amplitudes within a range $\Delta a = 0.06$.
Choosing this fiducial model weakens the priors on the long wavelength modes, particularly the average mode which is given infinite variance.   

\begin{table}[htdp]
\begin{tabular}{c|c|c|c|c|c}
   \hline\hline   
       & & \multicolumn{2}{c|}{standard prior} &
\multicolumn{2}{c}{weak prior} \\  \cline{3-6}
    & & SNLS3 & Union2.1 & SNLS3 & Union2.1 \\ \hline
    &\multicolumn{1}{c|} {prior} &$+1.2$  &$+0.44$ & $+0.68$& $+0.29$\\ 
    &\multicolumn{1}{c|} {SN} &$-1.7$  &$+0.24$ & $-2.2$& $+0.38$\\ 
    &\multicolumn{1}{c|} {RSD} &$-1.1$  &$-1.8$ & $-1.0$& $-1.8$\\ 
   $\Delta\chi^2$ &\multicolumn{1}{c|} {BAO} &$-1.4$  &$-1.2$ & $-1.8$& $-1.4$\\ 
    &\multicolumn{1}{c|} {others} &$-2.1$  &$-1.1$ & $-2.0$& $-1.1$\\ 
    &\multicolumn{1}{c|} {All} &$-5.1$  &$-3.4$ & $-6.3$& $-3.6$\\\hline
 $S/N$  &\multicolumn{1}{c|} {All} &$2.3$  &$1.8$ & $2.5$& $1.9$\\  \hline
    &\multicolumn{1}{c|} {$i=1$} &$0.06\pm0.2$  &$0.1\pm0.2$ & $0.06\pm0.2$& $0.2\pm0.2$\\ 
    &\multicolumn{1}{c|} {$i=2$} &$-0.3\pm0.3$  &$-0.3\pm0.3$ & $-0.2\pm0.3$& $-0.2\pm0.4$\\ 
  $\alpha_i$  &\multicolumn{1}{c|} {$i=3$} &$1.3\pm0.6$  &$0.9\pm0.6$ & $1.4\pm0.8$& $0.8\pm0.8$\\ 
    &\multicolumn{1}{c|} {$i=4$} &$-$  &$-$ & $0.6\pm0.4$& $0.5\pm0.4$\\ \hline\hline

  \end{tabular}
\caption{The $\Delta\chi^2$ rows: the improved $\chi^2$ of the best fit $w(z)$ model with respect to that of the best fit $\Lambda$CDM model, namely, 
$\chi^2_{w(z)}-\chi^2_{\Lambda {\rm CDM}}$. The $S/N$ row: the significance of $w(z)$ deviating from $-1$, which is simply $\sqrt{|\Delta\chi^2_{\rm tot}|}$. The $\alpha_i$ rows: the mean and 68\% CL error on the principal component amplitudes.}
\label{table}
\end{table}

We apply our method to a joint dataset of the latest cosmological observations including SN, CMB, LSS and the measurement of Hubble parameter at various redshifts. We compare two different SN samples: 
SNLS 3-year \cite{Conley:2011ku} and 
Union2.1 
\cite{Suzuki:2011hu}.
We used the 
WMAP 7-year CMB spectra 
\cite{Larson:2010gs}, the $H(z)$ data compiled by \cite{Moresco:2012by}, the $f\sigma_8$ estimates
from the redshift space distortions (RSD) 
and the baryonic acoustic oscillations (BAO) measurements from SDSS-II \cite{Reid:2009xm}, SDSS-III BOSS \cite{boss}, 6dF \cite{6df} and the WiggleZ survey \cite{wigglez}. Given this joint dataset, we use MCMC  \cite{Lewis:2002ah} to sample the parameter space ${\bf P} \equiv (\omega_{b}, \omega_{c}, \Theta_{s}, \tau, n_s, A_s, w_1, ..., w_{20},\mathcal{N})$ where $\omega_{b}$ and $\omega_{c}$ are the baryon and cold dark matter densities, $\Theta_{s}$ is the ratio of the sound horizon to the angular diameter distance at decoupling, $\tau$ is the optical depth, $n_s$ and $A_s$ are the primordial power spectral index and amplitude and $w_1,...,w_{20}$ denote the 20 $w$ bins uniform in $a$ from $a_{\rm min}=0.4$ to $a=1$, corresponding to $z\in[0,1.5]$. We fix $w=-1$ at $z>1.5$ \footnote{
Adding a $w$ bin from $z=1.5$ to $1100$ has a negligible effect on our 
results.}. 
We also 
marginalize over 
parameters 
accounting for the calibration uncertainty in measuring the intrinsic SN luminosity. We use a modified version of {\tt CAMB} \cite{CAMB} to calculate the observables and include dark energy perturbations for an arbitrary ${\bf w}$ following the prescription in \cite{DEP}.

\begin{figure*}[tbp]
\includegraphics[scale=0.36]{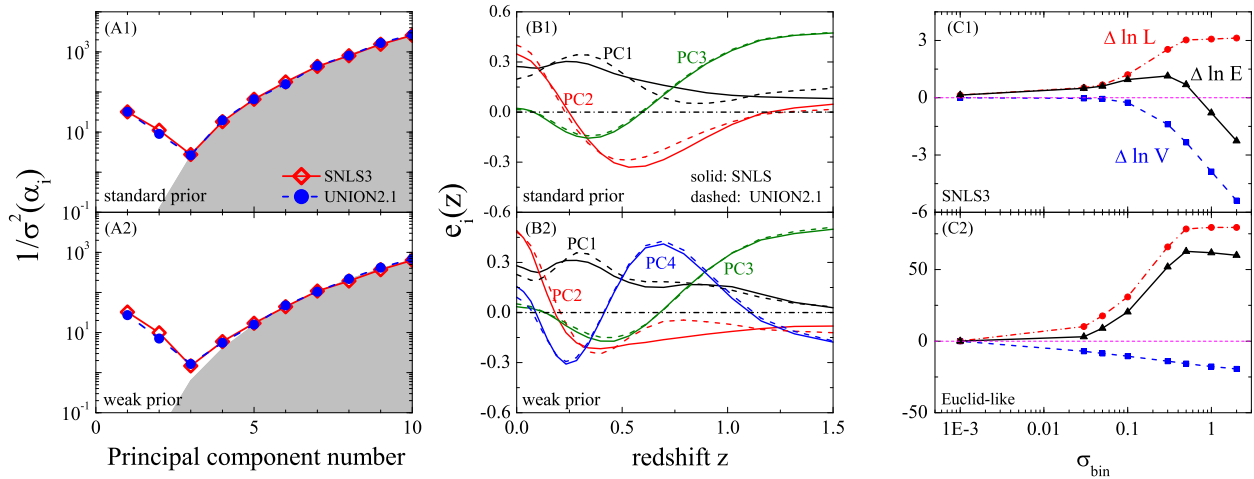}
\caption{Panels (A1,A2): Eigenvalues of the covariance matrix of the $w$ bins obtained from the data plus prior (diamonds and circles) and prior alone 
(shaded region) using MCMC. Panels (B1,B2): The first 3 (4) data-dominated eigenmodes of the covariance matrices. These panels show only small differences arise when different SN datasets (SNLS3 and Union2.1) are combined with other data. Results for the standard prior (A1,B1) and weak prior (A2,B2) are shown separately. Panels C1 and C2 show the improved likelihood ($\Delta$ln$L$), the reduced fraction of the sampled parameter space ($\Delta$ln$V$), and \tr{the logarithm of the} evidence ratio as a function of $\sigma_{\rm bin}$ relative to the $\Lambda$CDM model (which is the small $\sigma_{\rm bin}$ limit.) C1 and C2 are for the present data (SNLS3) and the simulated Planck and Euclid-like data respectively. }\label{fig:evalue}
\end{figure*}
  
Fig. \ref{fig:wz} shows the reconstruction results obtained with MCMC sampling of the parameter space ${\bf P}$ to find the minimum of $\chi^2_{\rm data} + \chi^2_{\rm prior}$. The shaded regions in panels (A1-A4) illustrate the 68 and 95\% CL uncertainty of the reconstruction, while the solid curves show the best fit $w(z)$ models. We show separately the results using standard or weak priors, and using the SN data of SNLS3 or Union2.1, combined with all the other data.  Looking at the different priors, we see similar trends, but the reconstruction uncertainties for the weaker prior are larger at low $z (z\simeq0)$ and high $z (z\gtrsim0.6)$, where the constraints from data are weak. At the sweet spot, $z\simeq0.15$, the error ($\Delta w|_{z=0.15}\simeq0.12$) is independent of the prior, since there the data are sufficiently strong to outweigh the priors. Interestingly, the most likely reconstruction using the SNLS3 data favors a transition from $w<-1$ at low redshift to $w>-1$ at higher redshift, a behavior that is consistent with the quintom model which allows $w$ to cross $-1$ \cite{Feng:2004ad}. The result for Union2.1 shows a similar trend, but at a lower significance. As shown in Table~\ref{table}, for SNLS3 combined with all other data, the significance of $w$ departing from $-1$ is 2.3(2.5)-$\sigma$ under the standard(weak) prior.

To understand this result, we show the reconstruction ranges using different data combinations in panels (B1-B4). Without the RSD and BAO data (black solid), SNLS3 data slightly prefers a model with $w<-1$ at $z\sim0.25$, but Union2.1 is very consistent with $w=-1$. With the RSD data (magenta dashed), the reconstruction at $z\gtrsim0.25$ is better constrained, and $w(z)$ is slightly increased at $z>0.25$. After adding BAO, the evolution from $w< -1$ at low redshifts to $w>-1$ at high redshifts becomes more evident, particularly when using the SNLS3 data; this suggests that the BAO data is at least partially responsible for the trend.  As shown in panel C1 for the SNLS3 data and a weak prior, the best fit $w(z)$ can fit the BAO data (the comoving angular diameter distance $D_V$) better than the $\Lambda$CDM model, especially for the BOSS BAO data point at $z\sim0.57$ (the $\chi^2$ for BAO is reduced by 1.8 as shown in Table \ref{table}). This naturally explains the result we found: $w$ needs to be less than $-1$ at low redshifts to get a higher $D_V$ at $z\sim0.6$ compared to $\Lambda$CDM,
while CMB requires the average $w$ to be 
close to $-1$, so the high redshift behaviour is pushed above $-1$ to compensate. This quintom behavior is also favored by SNLS3, where the contribution to $\chi^2$ from SN is reduced by 2.2. 

We also perform a PCA on the posterior distribution, identifying the uncorrelated collective $w(z)$ degrees of freedom after marginalizing over the other parameters.  In Fig, \ref{fig:evalue}, panels (A1,A2) show the eigenvalues of data plus prior (points) and prior alone (shaded regions).  The choice of the standard (weak) prior allows \tr{ the information in
the strongest 2(3) data modes to survive the addition of the prior}. In panels (B1,B2), we show the data-dominated eigenmodes $e_i(z)$.  We can project $w(z)$ onto this orthonormal basis, \ie~ $1+w(z)=\sum_i \alpha_i e_i(z).$ Note that if $w=-1$, all the $\alpha's$ should be zero.
Projecting the best fit $w(z)$ models onto the eigenbasis, we obtain the constraint (central value and 68\% CL error) on the $\alpha$'s shown in Table \ref{table}. \tr{We can see that $\Delta\chi^2\simeq\sum_{i=1}^{4}[\alpha_i/\sigma(\alpha_i)]^2$, meaning that the improvement in the fits is almost entirely accounted for by these three or four principal components.}

The reconstructed models necessarily provide a better fit than $\Lambda$CDM (Table \ref{table}), but the crucial issue is whether the improvement in fit is enough to compensate for the increased parameter space in the prior model.  Naively, the non-parametric binning has 20 additional degrees of freedom, but this far overestimates the true parameter volume when the correlated prior is imposed; in the limit of a very strong prior, the model can become equivalent to the cosmological constant model itself.  

To quantify this, we estimate the evidence ratio, called the Bayes' factor, within a family of models which interpolates smoothly between the prior described above and $\Lambda$CDM; this is implemented by adding a larger and larger diagonal term to the prior which shrinks the variance in each bin, $\chi^2_{\rm prior} = ({\bf w +1})^T[{\bf \tilde{C}}^{-1}+ \sigma_{\rm bin}^{-2}{\bf I}]({\bf w +1})$,  and shifts all the eigenvalues by a constant.  We calculate the evidence \cite{Heavens:2007ka}, $ E \equiv \int d^n {\bf P} \, {\cal P}({\bf D| P})\, {\cal P}_{\rm prior}({\bf P})$, by assuming a Gaussian prior for all parameters ${\bf P}$, described by a more general covariance matrix, ${\bf {\cal{ C}}}_{\rm prior}$, and also assume the resulting posterior is a Gaussian function of the parameters around the best fit, with covariance described by ${\bf {\cal{ C}}}_{\rm post}$.  We find: 
\be
E \propto \left(\frac{\det{{\cal C}_{\rm post}}}{\det{{\cal C}_{\rm prior}}}\right)^{1/2} e^{-\chi_{\rm b.f.}^2/2}, 
\ee 
where $\chi_{\rm b.f.}^2$ is the minimum total $\chi^2$ given the data and prior. 
The first term quantifies the fraction of the initial prior parameter volume which remains consistent with the measured data, while the second term accounts for how well the model is capable of fitting the data.   

We plot the \tr{logarithm of the Bayes' factor} in Fig. \ref{fig:evalue}, comparing a range of models approaching  $\Lambda$CDM to $\Lambda$CDM itself. 
For the original prior discussed above, marginalizing over fiducial models formally set the variance in the mean to infinity, meaning $\det{{\bf {\cal{ C}}}_{\rm prior}}$ diverges and the evidence ratio approaches zero.  However, choosing a more reasonable variance in the mean equation of state, one finds evidences comparable to $\Lambda$CDM. Perhaps surpisingly, for some choices of this variance, \tr{dynamical dark energy} is actually preferred over $\Lambda$CDM.  However, choosing a particular prior after seeing the evidence is \emph{post hoc}, and to be convinced one would need to see it for a range of priors most people would find reasonable.  

While current data is unable to distinguish $\Lambda$CDM from the best fit models, it is interesting to see what the future data could do. For this purpose, we perform a future forecast assuming that today's best fit model was the true model. We assume a future dataset of a deep SN survey \cite{Astier:2010qf} which is similar to the expected Euclid mission \cite{Euclid}, and the CMB data from Planck\cite{Planck}. The reconstruction can be performed using the Wiener filter projection \cite{Crittenden:2011aa} and the reconstruction is shown in Fig~\ref{fig:wz}. The yellow dots with error bars show $w(z)$, which would be reconstructed with very little bias. In both cases, the dynamical models can be clearly distinguished from $w=-1$.  This is supported by the evidence calculation in the lower right of Fig. \ref{fig:evalue}; for the Euclid-like data, the evidence ratio is large and sustained over a wide range of priors. 

Our non-parametric reconstruction 
is simple to implement and 
allows for straightforward interpretation and calculation of evidence ratios.  While we
see tantalizing hints for dynamical dark energy, the case is by no means conclusive.  
It is interesting that, 
except for 
Union2.1,
the fits to all the data are improved with our reconstructed model (Table \ref{table}).  
Note that while SNLS3 and Union2.1 overlap significantly, SNLS3 
has a more homogeneous higher redshift sample, which may make it less susceptible to systematic errors.  Future data will
be able to distinguish this reconstruction from $\Lambda$CDM,
while Planck and the full three-year SDSS SN sample might help in addressing this question in the near term. 

\acknowledgments We thank Chris D'Andrea and Will Percival for useful discussions.
GBZ and RC are supported by STFC grant ST/H002774/1, and LP by NSERC. XZ is supported in part by NSFC.


\begin{thebibliography}{99}

\bibitem{RiessPerl} A.~Riess {\etal}, Astron.~J.~{\bf 116}, 1009 (1998);
S.~Perlmutter {\em et al}, Ap.~J.~{\bf 517}, 565 (1999); For recent reviews, see   J.~Frieman, M.~Turner and D.~Huterer,
  Ann.\ Rev.\ Astron.\ Astrophys.\  {\bf 46}, 385 (2008) and 
D.~H.~Weinberg, M.~J.~Mortonson, D.~J.~Eisenstein, C.~Hirata, A.~G.~Riess and E.~Rozo,
  arXiv:1201.2434 [astro-ph.CO].


\bibitem{Sahni:2006pa}
  D.~Huterer, M.~S.~Turner,
  Phys.\ Rev.\  {\bf D60 } (1999)  081301; T.~D.~Saini, S.~Raychaudhury, V.~Sahni, A.~A.~Starobinsky,
  Phys.\ Rev.\ Lett.\  {\bf 85 } (2000)  1162-1165;   T.~Chiba, T.~Nakamura,
  Phys.\ Rev.\  {\bf D62 } (2000)  121301;  D.~Huterer, M.~S.~Turner,
  Phys.\ Rev.\  {\bf D64 } (2001)  123527; J.~Weller, A.~Albrecht,
  Phys.\ Rev.\  {\bf D65 } (2002)  103512; U.~Alam, V.~Sahni, T.~D.~Saini, A.~A.~Starobinsky,
  Mon.\ Not.\ Roy.\ Astron.\ Soc.\  {\bf 344 } (2003)  1057;  R.~A.~Daly, S.~G.~Djorgovski,  Astrophys.\ J.\  {\bf 597 } (2003)  9-20;  
  D.~Huterer and A.~Cooray,
  Phys.\ Rev.\ D {\bf 71}, 023506 (2005); 
A.~Shafieloo, U.~Alam, V.~Sahni, A.~A.~Starobinsky,
  Mon.\ Not.\ Roy.\ Astron.\ Soc.\  {\bf 366 } (2006)  1081-1095;  V.~Sahni and A.~Starobinsky,
  Int.\ J.\ Mod.\ Phys.\  D {\bf 15} (2006) 2105; J.~Dick, L.~Knox, M.~Chu,
  JCAP {\bf 0607 } (2006)  001; A.~Shafieloo,
  Mon.\ Not.\ Roy.\ Astron.\ Soc.\  {\bf 380 } (2007)  1573-1580;   A.~Albrecht and G.~Bernstein,
  Phys.\ Rev.\ D {\bf 75}, 103003 (2007);
  arXiv:0901.0721 [astro-ph.IM];   M.~J.~Mortonson, W.~Hu and D.~Huterer,
  Phys.\ Rev.\ D {\bf 79}, 023004 (2009);   M.~J.~Mortonson, D.~Huterer and W.~Hu,
  Phys.\ Rev.\ D {\bf 82}, 063004 (2010);  J.~Tang, F.~B.~Abdalla and J.~Weller, Mon.\ Not.\ Roy.\ Astron.\ Soc.\, {\bf 416} (2011) 2212-2232. 

\bibitem{HutStark} D. Huterer \& G. Starkman, Phys. Rev. Lett. {\bf 90}, 031301 (2003).

\bibitem{Crittenden:2005wj}
  R.~G.~Crittenden, L.~Pogosian and G.~B.~Zhao,
  JCAP {\bf 0912} (2009) 025.

\bibitem{Clarkson:2010bm}
  C.~Clarkson and C.~Zunckel,
  Phys.\ Rev.\ Lett.\  {\bf 104} (2010) 211301.

\bibitem{Crittenden:2011aa}
  R.~G.~Crittenden, G.~B.~Zhao, L.~Pogosian, L.~Samushia and X.~Zhang,
  JCAP {\bf 1202} (2012) 048.

\bibitem{Holsclaw:2010sk}
  T.~Holsclaw {\it et al.},
  Phys.\ Rev.\ Lett.\  {\bf 105} (2010) 241302.
  
\bibitem{Shafieloo:2012ht}
  A.~Shafieloo, A.~G.~Kim and E.~V.~Linder,
  arXiv:1204.2272 [astro-ph.CO].

\bibitem{Seikel:2012uu} 
  M.~Seikel, C.~Clarkson and M.~Smith,
  JCAP {\bf 1206}, 036 (2012).


\bibitem{Conley:2011ku} 
  A.~Conley {\it et al.},
  Astrophys.\ J.\ Suppl.\  {\bf 192}, 1 (2011).

\bibitem{Suzuki:2011hu}
  N.~Suzuki {\it et al.},
  Astrophys.\ J.\  {\bf 746} (2012) 85.


\bibitem{Larson:2010gs}
  D.~Larson {\it et al.}  [WMAP Collaboration],
  Astrophys.\ J.\ Suppl.\  {\bf 192}, 16 (2011)

\bibitem{Moresco:2012by}
  M.~Moresco {\it et al.},
  arXiv:1201.6658 [astro-ph.CO].

\bibitem{Reid:2009xm}
  B.~A.~Reid {\it et al.},
  MNRAS {\bf 404} (2010) 60.

  
\bibitem{boss} 
  B.~A.~Reid {\it et al.},
  arXiv:1203.6641 [astro-ph.CO].

\bibitem{6df} 
  F.~Beutler {\it et al.},
 arXiv:1204.4725 [astro-ph.CO].

\bibitem{wigglez} 
  C.~Blake {\it et al.},
  MNRAS  {\bf 418}, 1707 (2011).


\bibitem{Lewis:2002ah}
  A.~Lewis and S.~Bridle,
  Phys.\ Rev.\  D {\bf 66} (2002) 103511.

\bibitem{CAMB} 

  A.~Lewis, A.~Challinor and A.~Lasenby,
  Astrophys.\ J.\  {\bf 538}, 473 (2000).  Available at \url{http://camb.info}

\bibitem{DEP} G.~B.~Zhao {\it et al.}, 
Phys.\ Rev.\  D {\bf 72}, 123515 (2005)  
  
\bibitem{Feng:2004ad}
  B.~Feng, X.~L.~Wang and X.~M.~Zhang,
  Phys.\ Lett.\  B {\bf 607}, 35 (2005).


\bibitem{Heavens:2007ka} 
  A.~F.~Heavens, T.~D.~Kitching and L.~Verde,
  Mon.\ Not.\ Roy.\ Astron.\ Soc.\  {\bf 380}, 1029 (2007).
  
\bibitem{Astier:2010qf}
  P.~Astier, J.~Guy, R.~Pain and C.~Balland, 2011, A\&A, 525, A7. 

\bibitem{Euclid} \url{http://sci.esa.int/euclid}

\bibitem{Planck} \url{http://www.rssd.esa.int/index.php?project=Planck}


\end{thebibliography}
\end{document}